\documentclass[11pt,a4paper]{article}
\usepackage{jinstpub}
\usepackage{mathrsfs}
\usepackage{bm}
\usepackage[separate-uncertainty=true,multi-part-units=single]{siunitx}
\usepackage[version=4]{mhchem}
\usepackage{lipsum}
\usepackage{booktabs}
\usepackage[caption=false]{subfig}

\begin{document}
\title{Optical property measurements of lithium chloride aqueous solution for a novel solar neutrino experiment}
\author[a,b,c]{Ye Liang,}
\author[a,b,c]{Tong Xu,}
\author[d]{Jialiang Zhang,}
\author[d]{Shuo Li,}
\author[d]{Ming Qi}
\author[a,b,c,*]{and Zhe Wang\note[*]{~Corresponding author.}}

\affiliation[a]{Department of Engineering Physics, Tsinghua University, Beijing 100084, China}
\affiliation[b]{Center for High Energy Physics, Tsinghua University, Beijing 100084, China}
\affiliation[c]{Key Laboratory of Particle \& Radiation Imaging (Tsinghua University), Ministry of Education, Beijing 100084, China}
\affiliation[d]{School of Physics, Nanjing University, Nanjing 210093, China}

\emailAdd{wangzhe-hep@tsinghua.edu.cn}

\arxivnumber{2211.05023}
\abstract{
The lithium chloride aqueous solution has great potential to be the detection medium of a novel solar neutrino detector. The nuclide \ce{^7 Li} provides a charged-current interaction channel with a high cross-section for the MeV-scale solar electron-neutrinos, enabling measurement of the solar neutrino spectrum. This work measures the optical properties and the light yields of a saturated lithium chloride solution. After adsorption with activated carbon and recrystallization, the solution shows little absorption in the sensitive wavelength range of the bialkali photomultipliers. The attenuation length is evaluated to reach 50 meters at 430 nm. 
In addition to being a pure Cherenkov detector medium, a wavelength shifter, carbostyril 124, is added to the LiCl aqueous solution. The compatibility and the enhancement of the light yield are confirmed, enabling the development of a water-based Cherenkov-enhanced lithium-rich detector.
}
\maketitle
\section{Introduction}

The measurements of solar neutrinos~\cite{davis1994review, ferrari2005gno, vignaud1998gallex, fukuda2003superkamiokande, boger2000sudbury} have triggered the study of neutrino mixing and oscillation and reveal the knowledge of the solar model. New experiments~\cite{lodovico2017hyperkamiokande, adam2015juno, beacom2017physics} are proposed to improve further the measurement of solar neutrino oscillation and solar model. 

The nuclide \ce{^7 Li} is an attractive candidate for the MeV-scale neutrino target~\cite{bahcall1989neutrino, haxton1996salty, alonso2014advanced, shao2022potential}. The reaction $\nu_e+\ce{^7 Li}\rightarrow e^- +\ce{^7 Be}$ provides a measurement of solar neutrino energy spectrum through a reaction of $E_{\nu_e}=E_T+\SI{0.862}{MeV}$. $E_T$ is the kinetic energy of the electron and the last value is the threshold of the neutrino charged-current (CC) channel on \ce{^7 Li}~\cite{bahcall1989neutrino}. The total CC cross section of \ce{^8 B} solar neutrino is \SI{3.5e-42}{cm^2}, nearly 60 times that of the neutrino elastic scattering on electrons~\cite{bahcall1989neutrino}. The energy of the emitted electron well reflects the energy of the incident neutrino. The direction of the electron is close to uniform, which can help distinguish the CC interactions from the elastic scatterings. 

Lithium chloride is highly soluble in water, \SI{82.8}{g/100~g} at \SI{20}{\celsius} (i.e.~45.29\% w/w, 7.5\% w/w of Li)~\cite{kwan2020crc}. The compound is cost-effective, neutral, and non-toxic, and its aqueous solution is readily accessible. The high concentration of lithium allows the design of a 10-meter scale solar neutrino detector, which has been revealed by a recent work~\cite{shao2022potential}. Previous experimental attempts of a LiCl detector, mentioned by Haxton~\cite{haxton1996salty}, seemed hindered by a lack of optical transparency, especially in the UV. However, the preparation and measurement details were not reported. Besides, lithium hydroxide (LiOH) was also suggested~\cite{haxton1996salty}, but this substance is strongly alkaline and is less water-soluble than LiCl (5\% w/w of LiOH solution reported, 1.5\% w/w of Li).

The addition of some water-soluble wavelength shifter is reported to enhance the light yield~\cite{dai2008wavelength}. The same technique can be applied to the LiCl solution to gain additional isotropic light yield comparable with the Cherenkov light. This shifted light has similar spatial characteristics to scintillation light and may further enhance the energy resolution without compromising the direction reconstruction~\cite{land2021mevscale,luo2023reconstruction}.

This work demonstrates that saturated LiCl solution is capable of being a Cherenkov detecting medium for a novel solar neutrino detector, and the possibility to enhance the energy resolution by adding a water-soluble wavelength shifter. The optical properties and the light yield of highly concentrated lithium chloride solutions are studied to provide evidence and parameters for detector design and simulation. Section~\ref{sec:prp} describes the preparation and purification of the LiCl solution and the addition of the wavelength shifter. Section~\ref{sec:OP} shows the absorption spectra of the purified solution and the measured attenuation length with a long tube. Section~\ref{sec:CY} measures the Cherenkov light yield of LiCl solution with a device using cosmic-ray muons and compares the solution with water. Section~\ref{sec:SY} shows the improved light yield of the water-based Cherenkov-enhanced lithium-rich detector when adding the wavelength shifter.

\section{Solution Preparation}\label{sec:prp}

LiCl salt is colorless, odorless, inflammable, and nontoxic; it is insensitive to acidity variation and is chemically stable in most cases. Industrial-grade LiCl contains ppm to ppb level impurities, which impair optical transparency. Microfiltration can conveniently remove insoluble and colloidal impurities, such as that in Ref.~\cite{yeh2010purification}. Adsorption with powdered activated carbon (PAC) and recrystallization are effective methods to improve the solution's transparency further. 

LiCl powder of 99\% purity is purchased from Shanghai Aladdin Biochemical Technology Co., Ltd. High purity deionized water with \SI{18.2}{\mega\ohm\per\cm} is produced with a pure water machine from Zhongyang company. All the glass containers are cleaned by ultrasonics with a detergent solution at \SI{60}{\celsius} for half an hour and then rinsed with a large quantity of pure water. To make a saturated solution, an excessive amount of LiCl is dissolved into pure water with a vigorous stir. The solution is then vacuum-filtered twice. The first filtration uses a piece of cellulose acetate membrane with a pole size of \SI{0.22}{\um}, and the second uses \SI{0.10}{\um}. After the filtrations, the solution becomes homogeneous, colorless, and transparent. 

PAC of 200-mesh, purchased from Shanghai Titan Scientific Co., Ltd., is used to remove impurities that may undermine UV transparency. Around \SI{2}{g} of PAC powder is added to every \SI{300}{mL} solution, stirred for 30 minutes, and filtrated using a \SI{0.10}{\um} membrane. The filtrate is then heated to boiling for recrystallization. Around a quarter of the water is removed by evaporation. After cooling to room temperature, the over-saturated lithium chloride crystallizes. The crystal is separated from the rest solution by microfiltration, dried by blowing clean air, and then dissolved into pure water to make up the product solution.

The saturated LiCl solution is measured to have a density of 1.27$\pm$\SI{0.01}{g/mL} and a concentration of 45.16$\pm$0.01\%w/w at \SI{20}{\celsius}. The density is obtained by measuring the mass of a \SI{10.00}{mL} sample of solution on an analytical balance; the concentration was obtained by evaporating a \SI{10.00}{mL} sample of solution at a temperature lower than \SI{150}{\celsius}, then cooling it to \SI{20}{\celsius} in a dry environment and weighing the residual solid mass. The errors are obtained by repeating the measurement three times.

Carbostyril 124 (C-124), a water-soluble wavelength shifter, is added to the solution to enhance the light yield. The substance of 97\% purity is purchased from Shanghai Bidepharm Technology Co., Ltd. It is a slightly gray powder with low solubility in water. One ppm C-124 is added to the purified saturated LiCl solution to make up the Cherenkov-enhanced lithium-rich water solution. The solution is stirred for several hours to have all the powder dissolved, and insoluble impurities are removed by filtration.

\section{Measurements of Optical Properties}
\label{sec:OP}

\subsection{Absorption Spectrum}
The absorption of LiCl is measured by a Lengguang Technology UV1910Pro double-beam spectrophotometer. A \SI{10}{cm} quartz cuvette is used, and the air is used as blank. The definition of absorbance is
\begin{equation}
  A = -\log_{10}\frac{I}{I_0},
\end{equation}
where $I_0$ is the intensity of the incident light, and $I$ is that of the transmitted light.

\begin{figure}[!htbp]
  \begin{center}
    \includegraphics[width=0.78\textwidth]{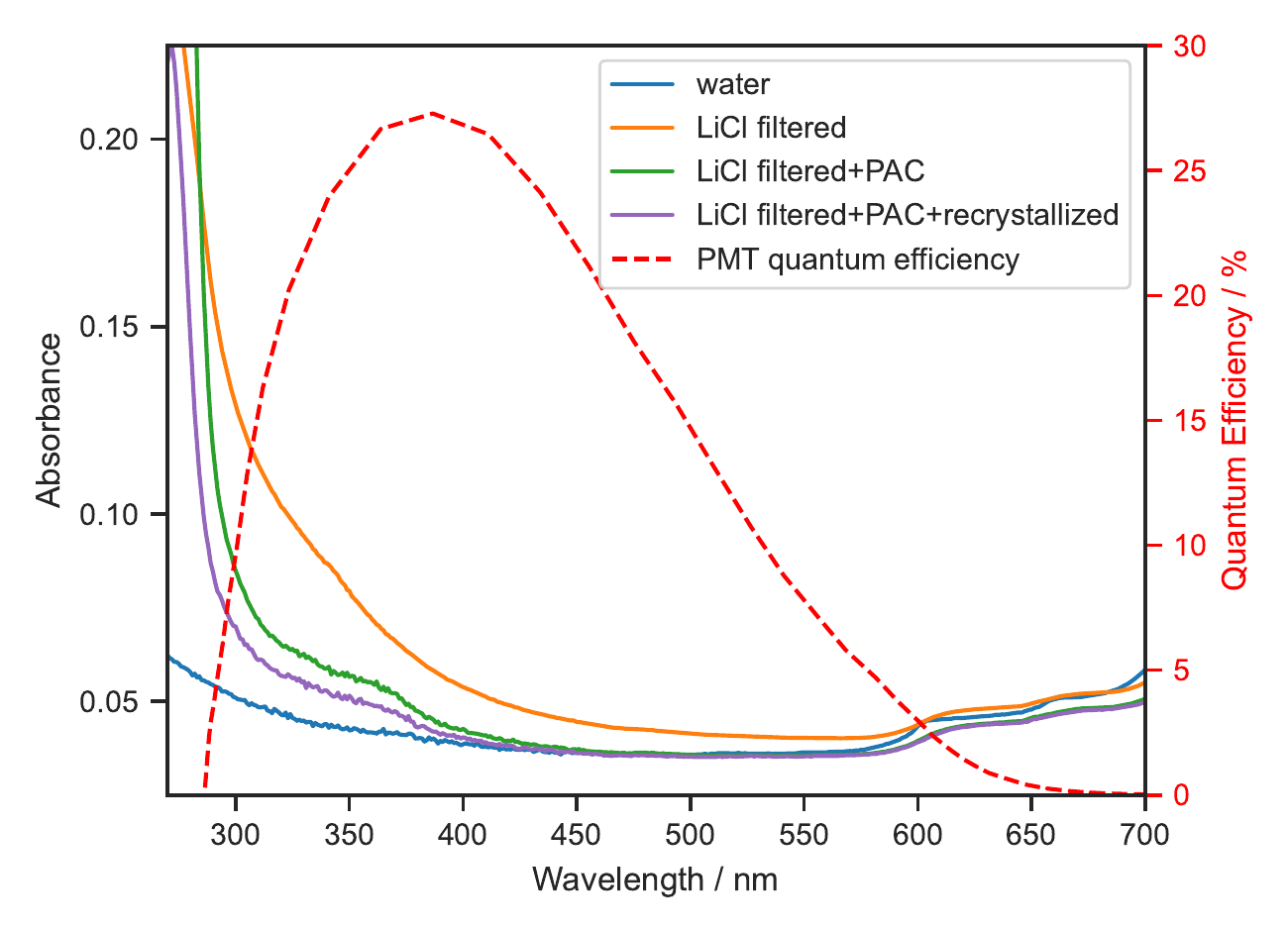}
    \caption{UV-vis absorption spectra of the saturated LiCl solution purified using different methods. The quantum efficiency spectrum of a typical bialkali PMT~\cite{hamamatsu} is shown as a reference.}\label{fig:uv_spectrum}
  \end{center}
\end{figure}

Figure~\ref{fig:uv_spectrum} shows the absorption spectra of saturated LiCl solution treated with filtration, PAC adsorption, and recrystallization. The saturated LiCl shows little absorption higher than water between \SI{400}{nm} to \SI{550}{nm}, which matches the sensitive region of typical bialkali photomultiplier tubes (PMTs) used in neutrino detectors. The slightly higher absorption than water is due to the impurities. PAC adsorption is shown effective in reducing the absorption of the solution, with an approximate 50\% decrease in absorbance near \SI{400}{nm}. A recrystallization step after the PAC process is shown to reduce the light absorption of the solution further. The absorbance around \SI{400}{nm} is very close to that of pure water.

\begin{figure}[!htbp]
  \begin{center}
    \subfloat[]{ \includegraphics[width=0.30\textwidth]{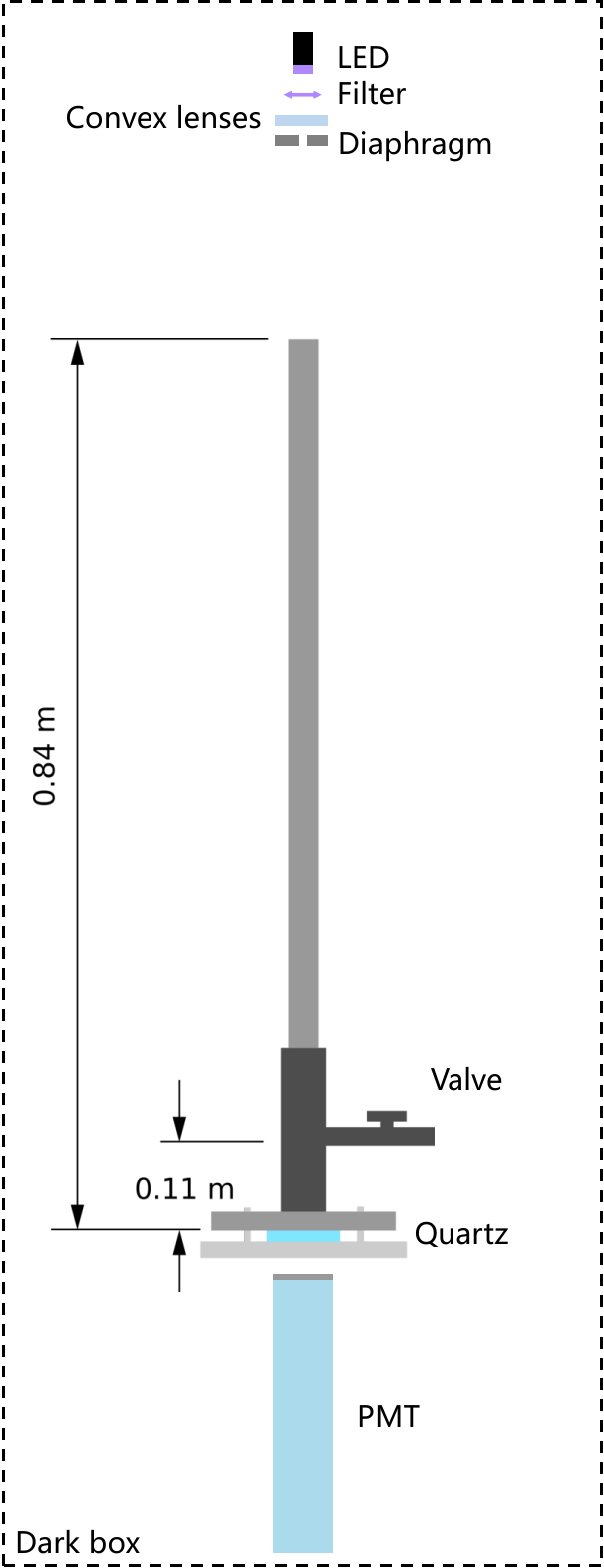} }
    \hspace{1.5cm}
    \subfloat[]{ \includegraphics[width=0.363\textwidth]{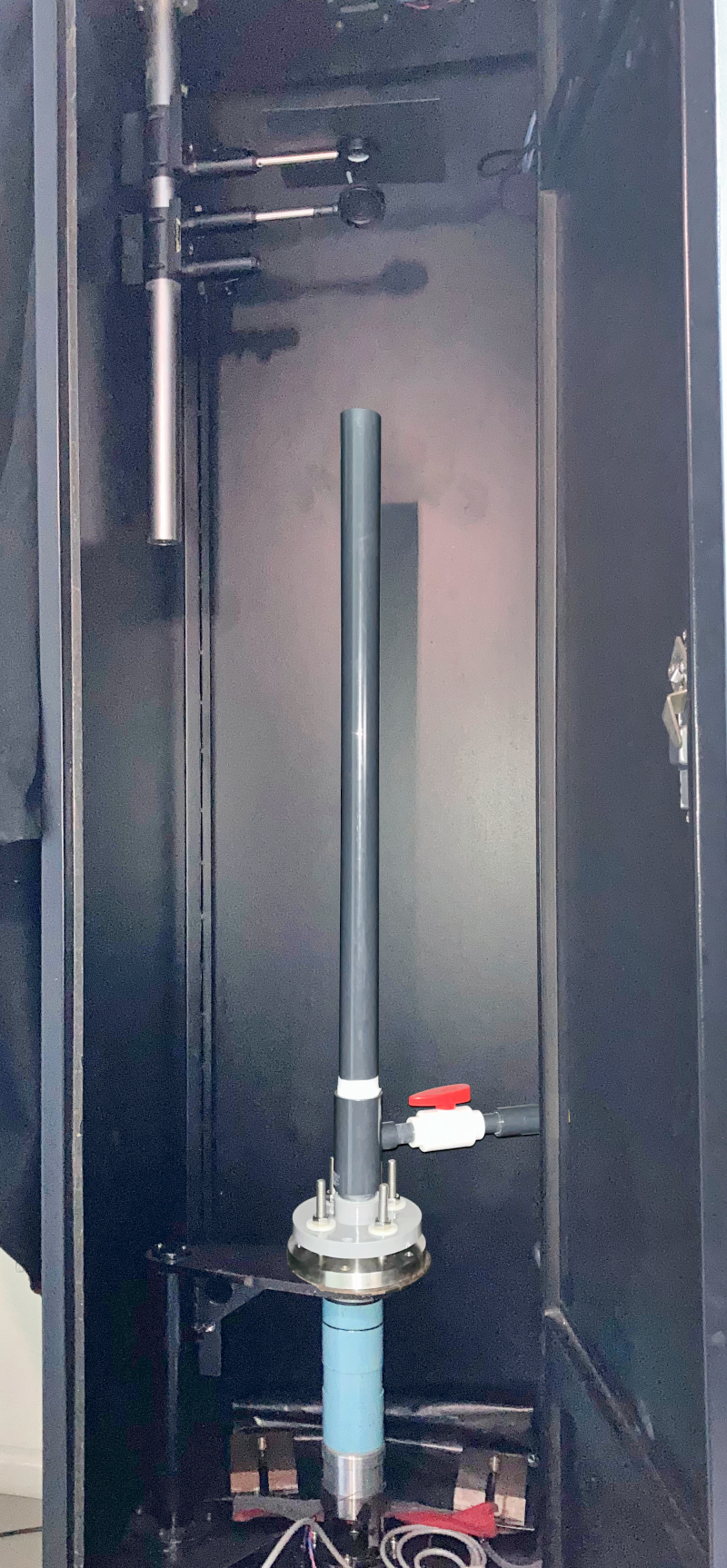} }
  \end{center}
    \caption{The tubular system for attenuation length measurement. (a) A sketch and (b) a photo of the system.}\label{fig:att}
\end{figure}

\subsection{Attenuation Length}

\begin{figure}[!htbp]
  \begin{center}
    \includegraphics[width=0.78\textwidth]{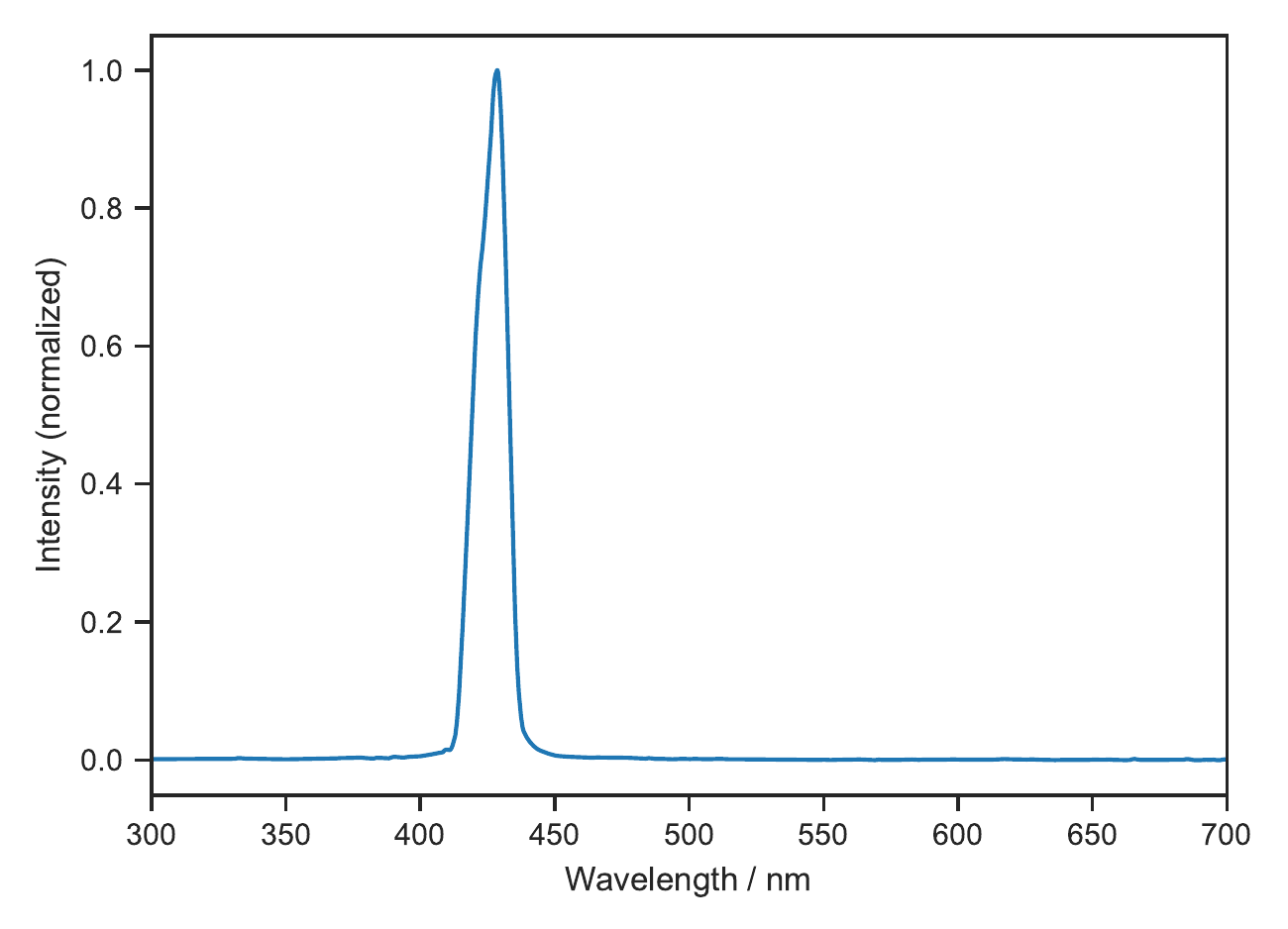}
    \caption{The spectrum of the LED light passing the \SI{430}{nm} filter in the tubular apparatus.}\label{fig:led}
  \end{center}
\end{figure}

The attenuation length of a saturated LiCl solution after careful purification is measured with a tubular system, shown in Fig.~\ref{fig:att}. The apparatus is built based on the attenuation length measurement system that deals with the liquid scintillators for Daya Bay and JUNO neutrino experiments~\cite{yu2016new,yang2017light,zhang2019using,cao2019light}. A black polyvinyl chloride (PVC) tube replaces the original stainless steel tube to prevent galvanic corrosion. The PVC tube is \SI{0.84}{m} long and has an inner radius of \SI{3.50}{cm}. The bottom of the tube is sealed with a thin piece of quartz transparent to UV light. The light source is a fiber-coupled light-emitting diode (LED), whose spectrum has a peak of \SI{430}{nm}. A Thorlabs FB430-10 optical filter is used to obtain monochromatic 430 nm light (FWHM=14, the spectrum shown in Fig.~\ref{fig:led}). The LED is controlled by a high-frequency signal generator. A pulse of light, emitted at a frequency of \SI{800}{Hz} and collimated by a group of lenses and a diaphragm, passes through the liquid in the tube and the quartz piece, and finally sheds onto a 2-inch Hamamatsu R7724 bialkali PMT. The light spot is around 1.5 mm in diameter and is aimed at the center of the photocathode. While sending pulse voltage to the LED, the signal generator triggers the data acquisition system simultaneously. The PMT signal within \SI{225}{ns} is read out by a CAEN V965 charge-to-digital converter~\cite{caen} and then converted to an analog-to-digital-conversion (ADC) value, which is proportional to the intensity of the transmitted light. The path length of the light through the liquid can be adjusted by the valve on the side. The whole apparatus is automatically controlled by a computer to avoid the error introduced by human operation. The entire instrument is housed in a dark box, which, alongside the electronic control system, is placed in a dark, grounded room. This nesting reduces stray light and shields against electromagnetic interference. The dark room is inside a thermostatic clean laboratory.

The measurement requires high stability of the apparatus; thus, a number of control operations and checks are performed. The environmental temperature is kept to fluctuate less than \SI{0.2}{\celsius} during the measurement. The liquid to be measured needs to be placed in the darkroom for two hours to reach the ambient temperature. After starting the electronics, it takes one hour for the temperature in the darkroom to stabilize. Then a standard water sample (Nongfu Spring, a commercially available drinking water) is first used to check the stability and calibrate the system. After filling the liquid and after every time adjusting the liquid level, the tube needs to be left for about ten minutes to eliminate the vibration and to make the liquid surface flat. The LED is adjusted to have sufficient intensity for sensitive measurement.

\begin{figure}[!htbp]
  \begin{center}
    \includegraphics[width=0.98\textwidth]{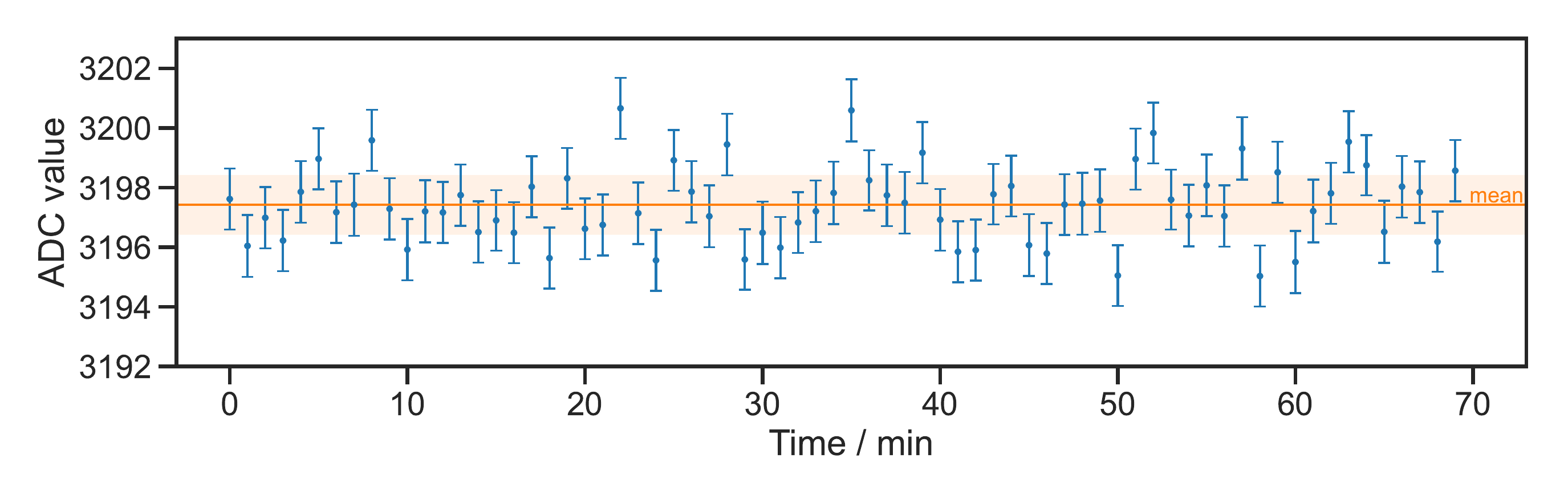}
    \caption{Fluctuation of the ADC values with 1$\sigma$ error bars at a specific liquid height over the measurement time scale.}\label{fig:stab}
  \end{center}
\end{figure}

In the measurement of the attenuation length of the LiCl solution, the path length varies from \SI{0.8}{m} to \SI{0.1}{m}, with a \SI{0.1}{m} decrement for 8 points. For each path length, five separate measurements are carried out. Each measurement persists for one minute to obtain around 20,000 triggered events, whose ADC values obey a Gaussian distribution; then a Gaussian fit is used to determine the mean value and the error. The statistic error is around 0.03\%. The stability of the measurement, shown in Fig.~\ref{fig:stab}, is checked by 70 consecutive measurements at a certain liquid height. It is shown that most of the data points deviate less than 1$\sigma$ from the mean value, showing good stability of the system.

\begin{figure}[!htbp]
  \begin{center}
  \subfloat[LiCl filtered and PAC adsorbed, batch 1]{ \includegraphics[width=0.5\textwidth]{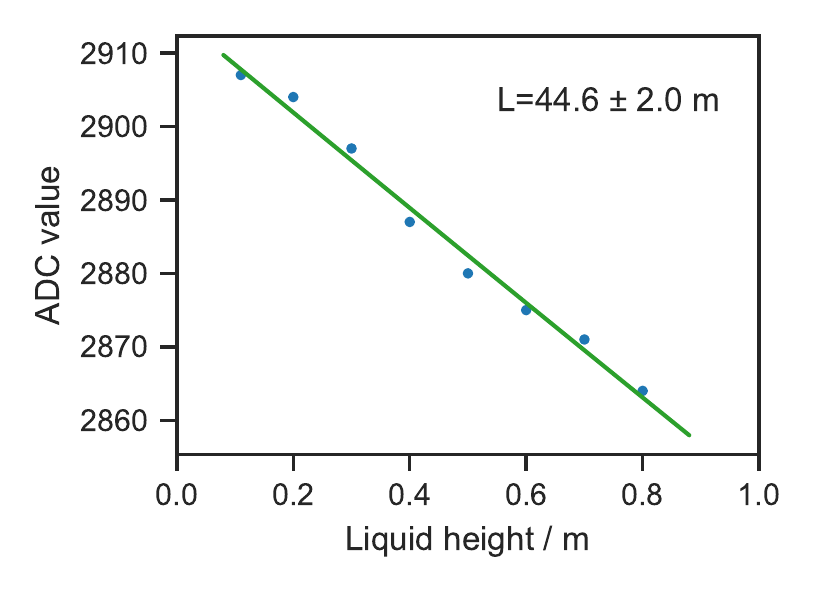} }
  \subfloat[LiCl filtered and PAC adsorbed, batch 2]{ \includegraphics[width=0.5\textwidth]{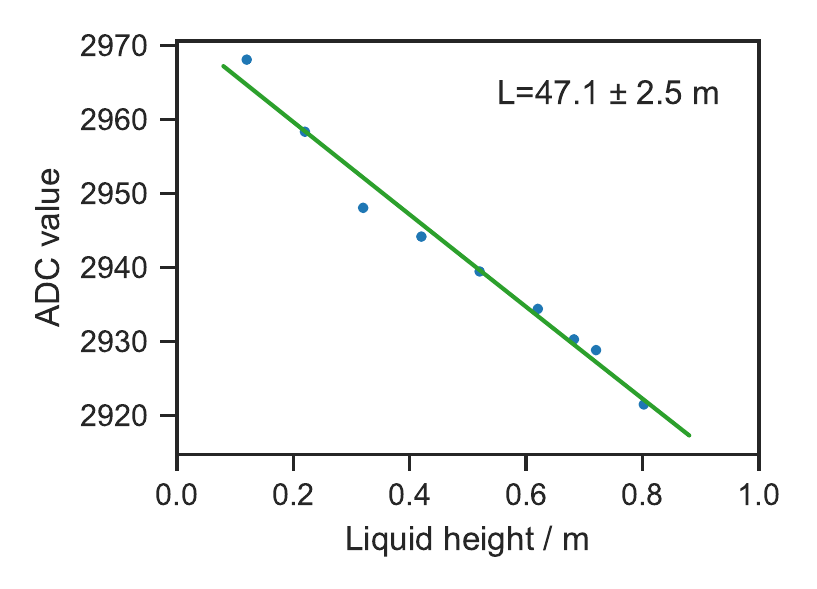} }\\
  \subfloat[LiCl filtered, PAC adsorbed, and recrystallized]{ \includegraphics[width=0.5\textwidth]{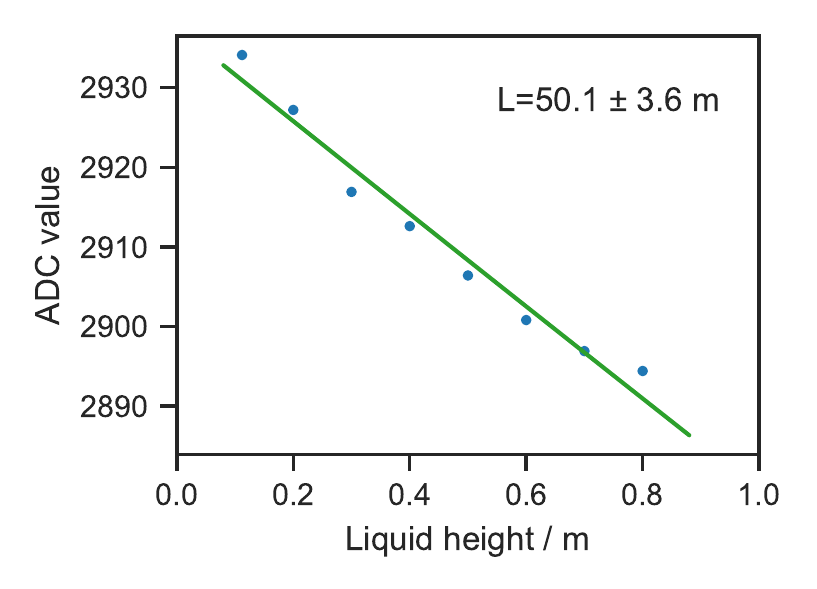} }\\
    \caption{Exponential fitting results of light attenuation in the saturated LiCl solutions. (a) and (b) are from two different batches processed by filtration and PAC adsorption, and (c) is additionally purified by recrystallization.}\label{fig:attfit}
  \end{center}
\end{figure}

The attenuation lengths of three saturated LiCl samples are measured. Two samples from different batches are purified using filtration and PAC adsorption, and the remaining one is further improved using recrystallization. The attenuation length of each sample is obtained by fitting an exponential function
\begin{equation}
  A = A_0 e^{-\frac{x}{L}},
\end{equation}
where $A$ is the ADC value, $x$ is the liquid height, and $L$ is the attenuation length. The pedestal caused by dark noise is deducted from ADC values in advance. Fitting results are shown in Fig.~\ref{fig:attfit}. All the LiCl samples are measured to have an attenuation length of more than \SI{40}{m}. The PAC adsorption process is shown to give stable purification results, \SI{44.6 +- 2.0}{m} and \SI{47.1 +- 2.5}{m}; the recrystallization process appears to be effective in further enhancing the attenuation length, bringing it to \SI{50.1 +- 3.6}{m}.

\section{Cherenkov Light Yield Verification}
\label{sec:CY}
\subsection{Experiment}

\begin{figure}[!htbp]
  \begin{center}
    \subfloat[]{ \includegraphics[width=0.5\textwidth]{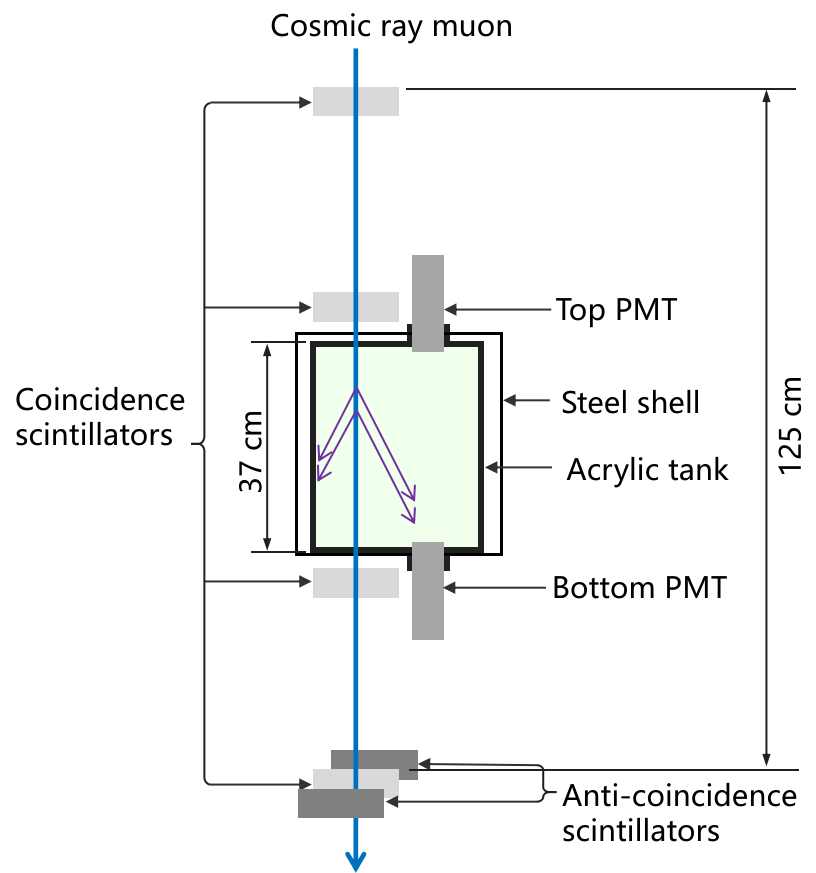} }
    \hspace{0.5cm}
    \subfloat[]{ \includegraphics[width=0.4\textwidth]{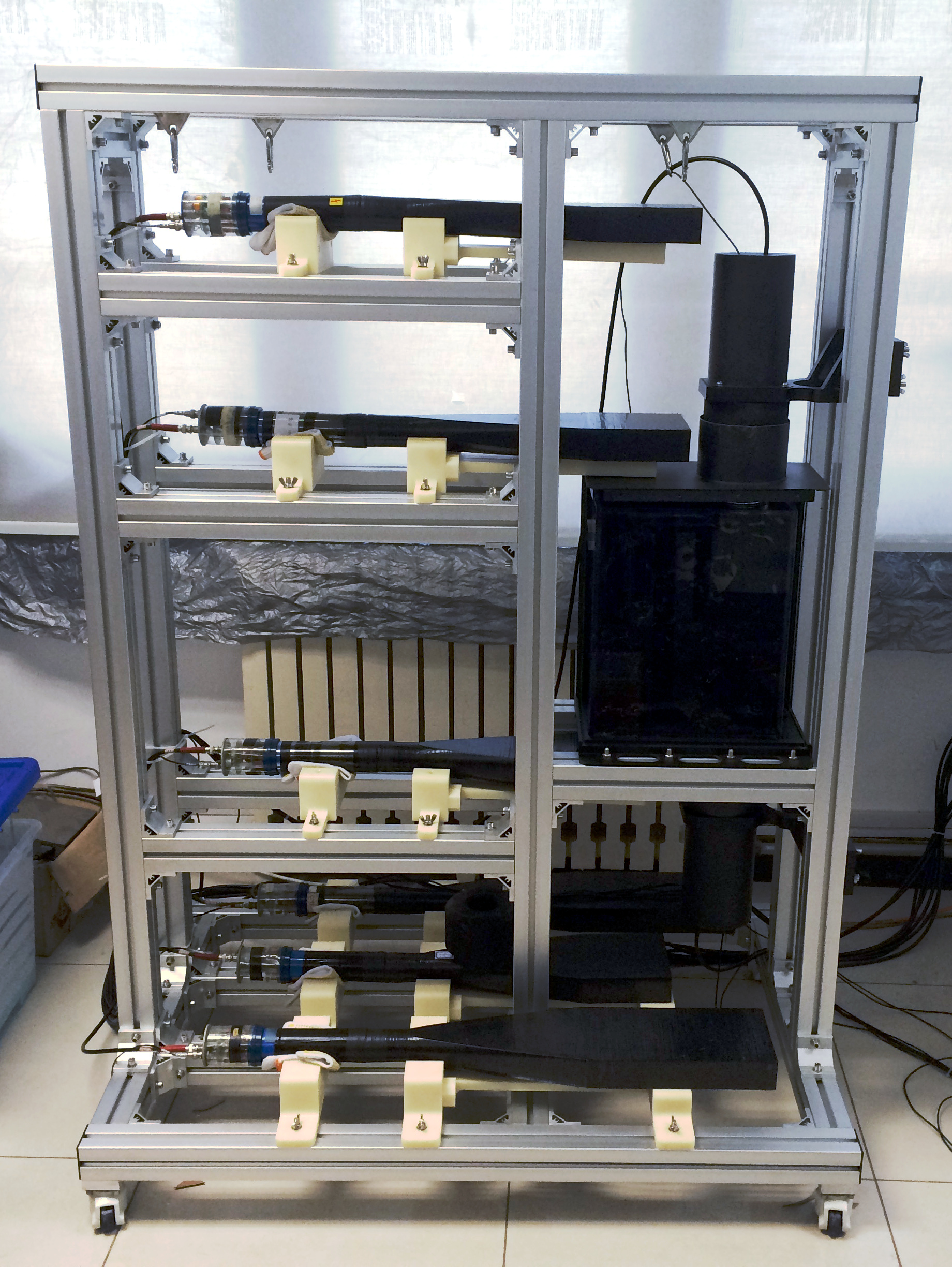} }
    \caption{The cosmic ray muon telescope for light yield measurements. (a) A sketch and (b) a photo.}\label{fig:20L}
  \end{center}
\end{figure}

The LiCl solution does not hinder the propagation of Cherenkov light, verified by measuring the Cherenkov light yield of the solution with a cosmic ray muon telescope (Fig.~\ref{fig:20L}). The device was used for the search of slow liquid scintillators~\cite{li2016separation,guo2019slow}. It comprises a \SI{15.4}{L} acrylic tank, four plastic coincidence scintillators, two anti-coincidence scintillators, and a pair of measuring PMTs. The tank contains the detecting medium; its inner surface is black and granulated to reduce the reflection. Four coincidence scintillators are used to trigger the data acquisition system when a cosmic ray muon is detected passing through the tank from top to bottom. The anti-coincidence scintillators on the side are used to tag the shower events which will be rejected later in the offline analysis. The measuring PMTs are two 2-inch Hamamatsu R1828-01 bialkali PMTs, mounted on the top and bottom of the detector tank. Their transit time spread (FWHM) is \SI{550}{ps}, and their gain is \num{2.0e7}. Their glass heads are slightly immersed in the liquid medium to avoid additional reflection in air layers. When a muon downwardly shoots into the medium, the bottom PMT measures the Cherenkov signals, while the top PMT detects the isotropic scintillation, if any. 
The whole device is well-covered to avoid ambient light interference.
A CAEN VX1721 flash analog-to-digital converter~\cite{caen} is used to process the trigger and read out the waveforms. 
The peak height and FWHM of each waveform are measured. 
The charge $Q$ of the waveform is calculated by integrating within the time interval from \SI{20}{ns} before the peak to \SI{60}{ns} after the peak.

Predominant noises are electronic noises and multi-muon events. The former are removed by checking the peak-to-charge ratio and peak-to-width ratio of their waveforms. The latter are removed by first checking if any signal is presented in the anti-coincidence channels, and then checking if the integrated charges in the coincidence channels deviate from the expected Landau distribution by a muon in a 5-cm plastic scintillator. Around 1,000 events are obtained in the measurement for each sample. The average waveforms of the top and the bottom PMT are shown in Fig.~\ref{fig:cherenkov}~(a). Then the  numbers of photoelectrons (PEs) of the top and bottom PMTs are calculated with
\begin{equation}
  N_{\text{PE}} = \frac{Q}{Q_\text{SPE}},
\end{equation}
where $Q_\text{SPE}$ is the gain of the PMT, which is calibrated using the single-photoelectron signals in dark noise. The major uncertainties are from PMT gain calibration and charge calculation and the total uncertainty is around 10\%. 

The Cherenkov PE yield of a purified saturated LiCl solution is measured. The Cherenkov PE yield of pure water is also measured for comparison. The results are shown in Table \ref{tab:che}. It is demonstrated that the solution does not hinder the propagation of Cherenkov light
and minor reflection is seen on the top PMT.

\begin{table}[h]
  \caption{Number of Photoelectrons (PEs) detected by measuring PMTs per cosmic-ray muon event.}\label{tab:che}
  \centering
  \begin{tabular}{lll}
\toprule
                 & Top PMT PEs    & Bottom PMT PEs \\ \midrule
Water            & 0.76$\pm$0.08  & 15.8$\pm$1.5  \\
Saturated LiCl solution            & 0.54$\pm$0.08 & 17.2$\pm$1.5  \\
Saturated LiCl solution with 1 ppm C-124 & 3.7$\pm$0.4   & 16.0$\pm$1.6  \\ \bottomrule
\end{tabular}
\end{table}

\subsection{Comparison with Simulation}
A Monte Carlo simulation based on GEANT4~\cite{agostinelli2003geant4,allison2006geant4,allison2016recent} is implemented to verify the experiment results. In the simulation, the detector tank with water as in the experiment is constructed and the top and the bottom PMT are placed. The energy of the incident muon is sampled with a power law distribution with an order of $-2.7$, which is the low energy asymptote of the Gaisser formula~\cite{gaisser2016cosmic}. Standard electromagnetic and muon-nucleus processes are included. The refraction indices of water for different wavelengths are set according to Ref.~\cite{daimon2007measurement}. Around 7,000 muon events with Cherenkov signals are obtained in the simulation. The result of the simulation is that the top PMT receives 0.27 PEs and the bottom PMT receives 15.6 PEs.

\section{Adding a Wavelength Shifter}\label{sec:SY}

For a detector with a diameter larger than ten meters, short-wavelength Cherenkov light in the UV range will be absorbed quickly and will not be detected. But this light can be converted to a longer PMT-sensitive wavelength to increase the overall light yield.

\begin{figure}[!htbp]
  \begin{center}
    \includegraphics[width=0.78\textwidth]{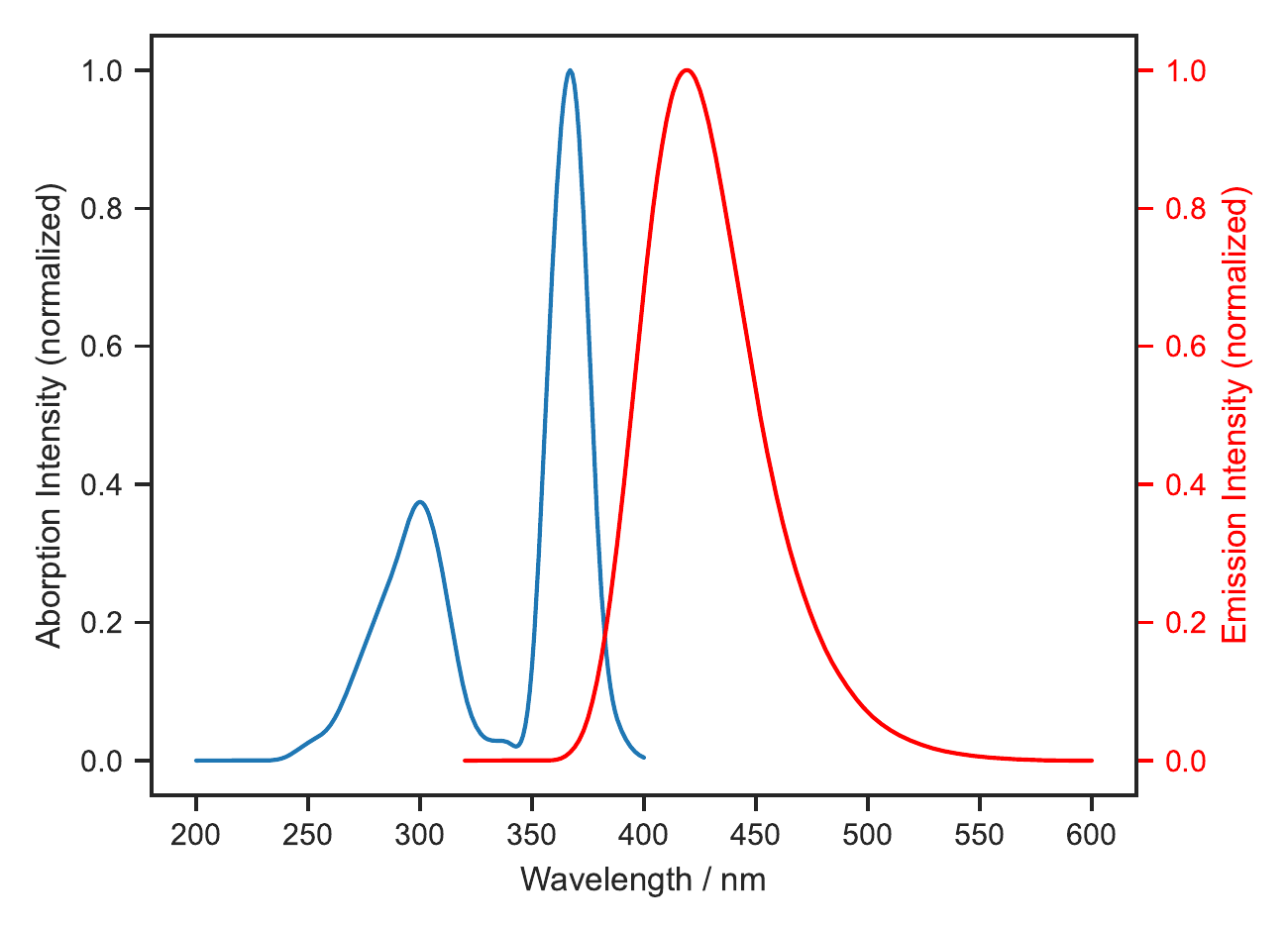}
    \caption{The absorption and emission spectrum of C-124 dissolved in water at a concentration of 62.5 ppm.}\label{fig:c124}
  \end{center}
\end{figure}

A wavelength shifter, carbostyril 124 (C-124), is added to the LiCl solution to enhance the detector performance. This compound has been investigated to increase the light yield of a pure water Cherenkov detector~\cite{dai2008wavelength}. The attenuation length of a ppm-level C-124 aqueous solution has been reported to be over \SI{40}{m} at 430 nm~\cite{so2014development}. The absorption and fluorescence emission spectrum of C-124 is measured again using a Lengguang Technology F97Pro fluorescence spectrophotometer. As shown in Fig.~\ref{fig:c124}, C-124 has strong light absorption below \SI{400}{nm} and emission in the region of 400–500 nm.  A solubility test shows that C-124 dissolves at least \SI{200}{ppm} in a saturated lithium chloride solution. A saturated LiCl solution with \SI{1}{ppm} C-124 is prepared, which has no visible changes when sealed and kept at room temperature for over two months.

\begin{figure}[!htbp]
  \begin{center}
    \subfloat[LiCl]{ \includegraphics[width=0.49\textwidth]{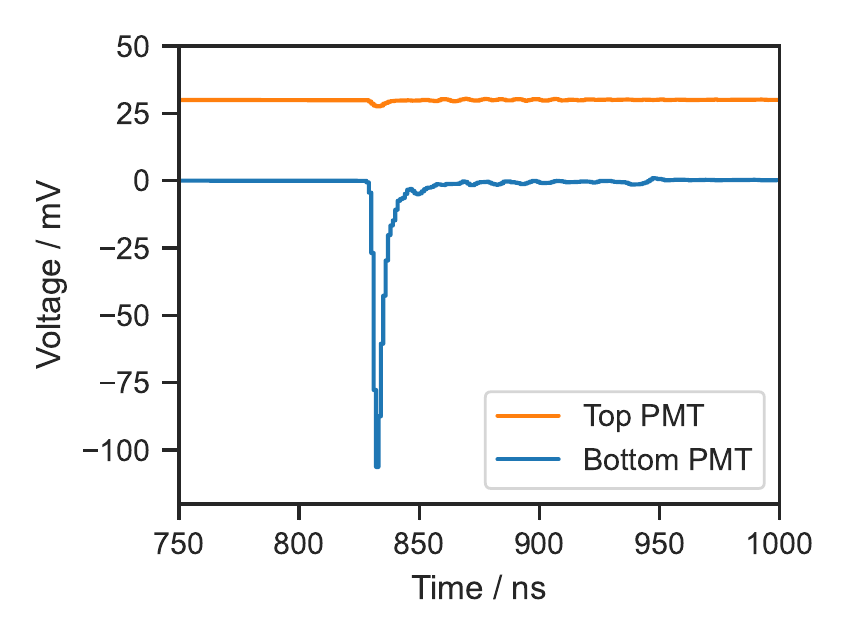} }
    \subfloat[LiCl with 1 ppm C-124]{ \includegraphics[width=0.49\textwidth]{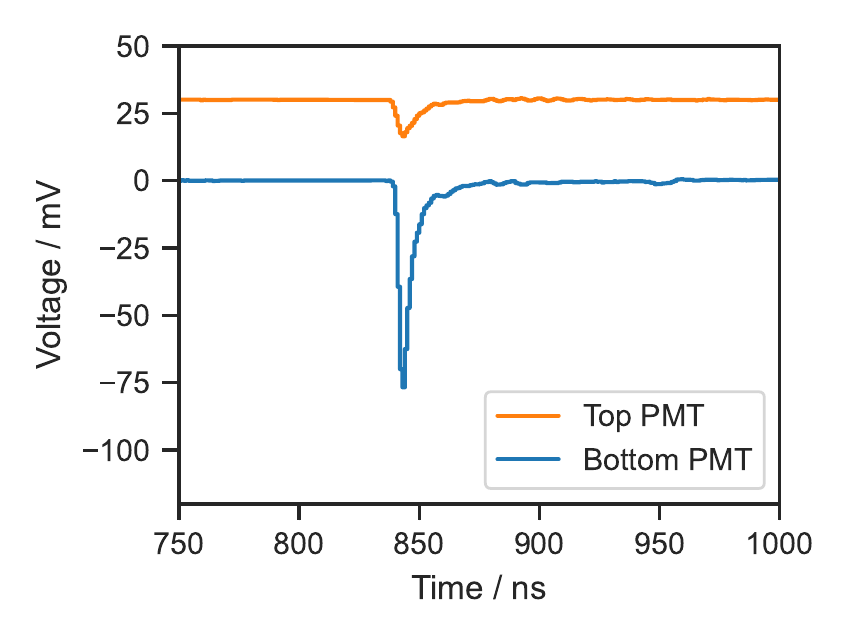} }
    \caption{The average waveforms from the top and the bottom PMT of the muon telescope for the LiCl solution.}\label{fig:cherenkov}
  \end{center}
\end{figure}

The Cherenkov and shifted light yield of the saturated LiCl solution with \SI{1}{ppm} C-124 is measured using the muon telescope described in the previous section. The average waveforms of the events after selection are shown in Fig.~\ref{fig:cherenkov}~(b). Comparing with the pure LiCl aqueous solution case in Fig.~\ref{fig:cherenkov}~(a), the top PMT receives more photon signals from isotropic emission. The result of measurement is listed in Table~\ref{tab:che}. 
The number of PEs of the bottom PMT is almost unchanged. The number of Cherenkov PEs is $16.0-3.7=12.3$, maintaining enough amount of Cherenkov light detection. 
Considering the 4$\pi$ emission feature of the shifted light, the total amount of shifted light is comparable to the Cherenkov light. 
As is explained in Ref.~\cite{luo2023reconstruction}, this solution can be used as a Cherenkov-enhanced detection medium, giving a reasonable direction resolution and better energy and vertex resolution than pure water.

\section{Summary and Conclusion}\label{sec:sum}

A Cherenkov detector or a Cherenkov-enhanced detector with concentrated lithium salt is attractive for MeV-scale solar neutrino detection. This work measures the optical properties of a saturated LiCl solution. After purification using PAC adsorption and recrystallization methods, the absorption spectrum of the LiCl aqueous solution is close to that of pure water in the sensitive range of common bialkali PMTs, and 
the attenuation length is measured to reach \SI{50}{m} at 430 nm. The Cherenkov light yield of the solution is measured using a cosmic ray muon telescope, which shows that the Cherenkov light induced by cosmic ray muons is consistent with pure water. Additionally, a ppm-level wavelength shifter C-124 is added to the saturated LiCl solution and a small amount of isotropic shifted light is observed along with the major Cherenkov light. The LiCl aqueous solution with C-124 can be used as a water-based Cherenkov-enhanced detector.

\section{Acknowledgement}
This work is supported in part by the National Natural Science Foundation of China (Nos.~12141503 and 11620101004), the Ministry of Science and Technology of China (No.~2022YFA1604704), the Key Laboratory of Particle \& Radiation Imaging (Tsinghua University), and the CAS Center for Excellence in Particle Physics (CCEPP).

\bibliographystyle{JHEP}
\bibliography{Library.bib}

\providecommand{\href}[2]{#2}\begingroup\raggedright\begin{thebibliography}{10}

\bibitem{davis1994review}
R.~Davis, \emph{A review of the homestake solar neutrino experiment},
  \href{https://doi.org/10.1016/0146-6410(94)90004-3}{\emph{Progress in
  Particle and Nuclear Physics} {\bfseries 32} (1994) 13}.

\bibitem{ferrari2005gno}
N.~Ferrari, J.C.~Lanfranchi and L.~Pandola, \emph{The {{GNO}} experiment},
  \href{https://doi.org/10.1016/j.nuclphysbps.2005.01.225}{\emph{Nuclear
  Physics B - Proceedings Supplements} {\bfseries 143} (2005) 560}.

\bibitem{vignaud1998gallex}
D.~Vignaud, \emph{The {{GALLEX}} solar neutrino experiment},
  \href{https://doi.org/10.1016/S0920-5632(97)00498-2}{\emph{Nuclear Physics B
  - Proceedings Supplements} {\bfseries 60} (1998) 20}.

\bibitem{fukuda2003superkamiokande}
S.~Fukuda, Y.~Fukuda, T.~Hayakawa, E.~Ichihara, M.~Ishitsuka, Y.~Itow et~al.,
  \emph{The {{Super-Kamiokande}} detector},
  \href{https://doi.org/10.1016/S0168-9002(03)00425-X}{\emph{Nuclear
  Instruments and Methods in Physics Research Section A: Accelerators,
  Spectrometers, Detectors and Associated Equipment} {\bfseries 501} (2003)
  418}.

\bibitem{boger2000sudbury}
J.~Boger, R.L.~Hahn, J.K.~Rowley, A.L.~Carter, B.~Hollebone, D.~Kessler et~al.,
  \emph{The {{Sudbury Neutrino Observatory}}},
  \href{https://doi.org/10.1016/S0168-9002(99)01469-2}{\emph{Nuclear
  Instruments and Methods in Physics Research Section A: Accelerators,
  Spectrometers, Detectors and Associated Equipment} {\bfseries 449} (2000)
  172}.

\bibitem{lodovico2017hyperkamiokande}
F.D.~Lodovico and o.b.o.t.H.-K.~Collaboration, \emph{The {{Hyper-Kamiokande
  Experiment}}}, \href{https://doi.org/10.1088/1742-6596/888/1/012020}{\emph{J.
  Phys.: Conf. Ser.} {\bfseries 888} (2017) 012020}.

\bibitem{adam2015juno}
T.~Adam, F.~An, G.~An, Q.~An, N.~Anfimov, V.~Antonelli et~al., \emph{{{JUNO}}
  conceptual design report}, {\emph{ArXiv Prepr. ArXiv150807166} (2015) }
  [\href{https://arxiv.org/abs/1508.07166}{{\ttfamily 1508.07166}}].

\bibitem{beacom2017physics}
J.F.~Beacom, S.~Chen, J.~Cheng, S.N.~Doustimotlagh, Y.~Gao, G.~Gong et~al.,
  \emph{Physics prospects of the {{Jinping}} neutrino experiment},
  \href{https://doi.org/10.1088/1674-1137/41/2/023002}{\emph{Chinese Phys. C}
  {\bfseries 41} (2017) 023002}.

\bibitem{bahcall1989neutrino}
J.N.~Bahcall, \emph{Neutrino Astrophysics}, {Cambridge University Press}
  (1989).

\bibitem{haxton1996salty}
W.C.~Haxton, \emph{Salty {{Water Cerenkov Detectors}} for {{Solar Neutrinos}}},
  {\emph{Phys. Rev. Lett.} {\bfseries 76} (1996) 4}.

\bibitem{alonso2014advanced}
J.R.~Alonso, N.~Barros, M.~Bergevin, A.~Bernstein, L.~Bignell, E.~Blucher
  et~al., \emph{Advanced {{Scintillator Detector Concept}} ({{ASDC}}): {{A
  Concept Paper}} on the {{Physics Potential}} of {{Water-Based Liquid
  Scintillator}}}, {\emph{ArXiv14095864 Hep-Ex Physicsnucl-Ex Physicsphysics}
  (2014) } [\href{https://arxiv.org/abs/1409.5864}{{\ttfamily 1409.5864}}].

\bibitem{shao2022potential}
W.~Shao, W.~Xu, Y.~Liang, W.~Luo, T.~Xu, M.~Qi et~al., \emph{The {{Potential}}
  to {{Probe Solar Neutrino Physics}} with {{LiCl Water Solution}}},
  {\emph{ArXiv220301860 Hep-Ex Physicsphysics} (2022) }
  [\href{https://arxiv.org/abs/2203.01860}{{\ttfamily 2203.01860}}].

\bibitem{kwan2020crc}
J.~Kwan, \emph{{{CRC}} Handbook of Chemistry and Physics}, {CRC Press} (2020).

\bibitem{dai2008wavelength}
X.~Dai, E.~Rollin, A.~Bellerive, C.~Hargrove, D.~Sinclair, C.~Mifflin et~al.,
  \emph{Wavelength shifters for water {{Cherenkov}} detectors},
  \href{https://doi.org/10.1016/j.nima.2008.01.101}{\emph{Nuclear Instruments
  and Methods in Physics Research Section A: Accelerators, Spectrometers,
  Detectors and Associated Equipment} {\bfseries 589} (2008) 290}.

\bibitem{land2021mevscale}
B.J.~Land, Z.~Bagdasarian, J.~Caravaca, M.~Smiley, M.~Yeh and G.D.~Orebi~Gann,
  \emph{{{MeV-scale}} performance of water-based and pure liquid scintillator
  detectors}, \href{https://doi.org/10.1103/PhysRevD.103.052004}{\emph{Phys.
  Rev. D} {\bfseries 103} (2021) 052004}.

\bibitem{luo2023reconstruction}
W.~Luo, Q.~Liu, Y.~Zheng, Z.~Wang and S.~Chen, \emph{Reconstruction algorithm
  for a novel {{Cherenkov}} scintillation detector},
  \href{https://doi.org/10.1088/1748-0221/18/02/P02004}{\emph{J. Inst.}
  {\bfseries 18} (2023) P02004}.

\bibitem{yeh2010purification}
M.~Yeh, J.B.~Cumming, S.~Hans and R.L.~Hahn, \emph{Purification of lanthanides
  for large neutrino detectors: {{Thorium}} removal from gadolinium chloride},
  \href{https://doi.org/10.1016/j.nima.2010.02.124}{\emph{Nuclear Instruments
  and Methods in Physics Research Section A: Accelerators, Spectrometers,
  Detectors and Associated Equipment} {\bfseries 618} (2010) 124}.

\bibitem{hamamatsu}
``Hamamatsu {{Photonics}}.'' https://www.hamamatsu.com/.

\bibitem{yu2016new}
G.-Y.~Yu, D.-W.~Cao, A.-Z.~Huang, L.~Yu, C.-W.~Loh, W.-W.~Wang et~al.,
  \emph{Some new progress on the light absorption properties of linear alkyl
  benzene solvent},
  \href{https://doi.org/10.1088/1674-1137/40/1/016002}{\emph{Chinese Phys. C}
  {\bfseries 40} (2016) 016002}.

\bibitem{yang2017light}
H.~Yang, D.~Cao, Z.~Qian, X.~Zhu, C.~Loh, A.~Huang et~al., \emph{Light
  attenuation length of high quality linear alkyl benzene as liquid
  scintillator solvent for the {{JUNO}} experiment},
  \href{https://doi.org/10.1088/1748-0221/12/11/T11004}{\emph{J. Inst.}
  {\bfseries 12} (2017) T11004}.

\bibitem{zhang2019using}
R.~Zhang, D.-W.~Cao, C.-W.~Loh, Y.-H.~Liu, F.-L.~Wu, J.-L.~Zhang et~al.,
  \emph{Using monochromatic light to measure attenuation length of liquid
  scintillator solvent {{LAB}}},
  \href{https://doi.org/10.1007/s41365-019-0542-1}{\emph{NUCL SCI TECH}
  {\bfseries 30} (2019) 30}.

\bibitem{cao2019light}
D.~Cao, R.~Zhang, Y.~Liu, C.~Loh, W.~Wang, Z.~Qian et~al., \emph{Light
  absorption properties of the high quality linear alkylbenzene for the
  {{JUNO}} experiment},
  \href{https://doi.org/10.1016/j.nima.2019.01.077}{\emph{Nuclear Instruments
  and Methods in Physics Research Section A: Accelerators, Spectrometers,
  Detectors and Associated Equipment} {\bfseries 927} (2019) 230}.

\bibitem{caen}
``{{CAEN}} - {{Tools}} for {{Discovery}}.'' https://www.caen.it/.

\bibitem{li2016separation}
M.~Li, Z.~Guo, M.~Yeh, Z.~Wang and S.~Chen, \emph{Separation of scintillation
  and {{Cherenkov}} lights in linear alkyl benzene},
  \href{https://doi.org/10.1016/j.nima.2016.05.132}{\emph{Nuclear Instruments
  and Methods in Physics Research Section A: Accelerators, Spectrometers,
  Detectors and Associated Equipment} {\bfseries 830} (2016) 303}.

\bibitem{guo2019slow}
Z.~Guo, M.~Yeh, R.~Zhang, D.-W.~Cao, M.~Qi, Z.~Wang et~al., \emph{Slow liquid
  scintillator candidates for {{MeV-scale}} neutrino experiments},
  \href{https://doi.org/10.1016/j.astropartphys.2019.02.001}{\emph{Astroparticle
  Physics} {\bfseries 109} (2019) 33}.

\bibitem{agostinelli2003geant4}
S.~Agostinelli, J.~Allison, K.~Amako, J.~Apostolakis, H.~Araujo, P.~Arce
  et~al., \emph{Geant4\textemdash a simulation toolkit},
  \href{https://doi.org/10.1016/S0168-9002(03)01368-8}{\emph{Nuclear
  Instruments and Methods in Physics Research Section A: Accelerators,
  Spectrometers, Detectors and Associated Equipment} {\bfseries 506} (2003)
  250}.

\bibitem{allison2006geant4}
J.~Allison, K.~Amako, J.~Apostolakis, H.~Araujo, P.~Arce~Dubois, M.~Asai
  et~al., \emph{Geant4 developments and applications},
  \href{https://doi.org/10.1109/TNS.2006.869826}{\emph{IEEE Trans. Nucl. Sci.}
  {\bfseries 53} (2006) 270}.

\bibitem{allison2016recent}
J.~Allison, K.~Amako, J.~Apostolakis, P.~Arce, M.~Asai, T.~Aso et~al.,
  \emph{Recent developments in {{Geant4}}},
  \href{https://doi.org/10.1016/j.nima.2016.06.125}{\emph{Nuclear Instruments
  and Methods in Physics Research Section A: Accelerators, Spectrometers,
  Detectors and Associated Equipment} {\bfseries 835} (2016) 186}.

\bibitem{gaisser2016cosmic}
T.K.~Gaisser, R.~Engel and E.~Resconi, \emph{Cosmic Rays and Particle Physics},
  {Cambridge University Press} (2016).

\bibitem{daimon2007measurement}
M.~Daimon and A.~Masumura, \emph{Measurement of the refractive index of
  distilled water from the near-infrared region to the ultraviolet region},
  \href{https://doi.org/10.1364/AO.46.003811}{\emph{Appl. Opt., AO} {\bfseries
  46} (2007) 3811}.

\bibitem{so2014development}
S.H.~So, K.K.~Joo, B.R.~Kim, B.K.~Kim, S.C.~Kim, C.D.~Shin et~al.,
  \emph{Development of a {{Liquid Scintillator Using Water}} for a {{Next
  Generation Neutrino Experiment}}},
  \href{https://doi.org/10.1155/2014/327184}{\emph{Adv. High Energy Phys.}
  {\bfseries 2014} (2014) e327184}.

\end{thebibliography}\endgroup
\end{document}